\documentclass[sigplan,screen]{acmart}
\usepackage{graphicx}
\usepackage{listings}
\usepackage{subcaption} 
\usepackage{multirow}
\usepackage{booktabs}
\usepackage{tcolorbox}
\usepackage{xspace}
\usepackage{xcolor}
\usepackage[ruled,vlined]{algorithm2e}
\usepackage{tabularx}
\usepackage{multicol}
\usepackage{array}
\usepackage{colortbl} 
\usepackage{multirow}
\usepackage{pbalance}
\usepackage{breakurl}
\usepackage{balance}
\usepackage{hyperref}
\hypersetup{breaklinks=true}
\newcommand{\toolname}{\textsc{GroupTuner}\xspace}

\AtBeginDocument{
  }

\setcopyright{acmlicensed}
\acmDOI{10.1145/3735452.3735530}
\acmYear{2025}
\copyrightyear{2025}
\acmISBN{979-8-4007-1921-9/25/06}
\acmConference[LCTES '25]{Proceedings of the 26th ACM SIGPLAN/SIGBED International Conference on Languages, Compilers, and Tools for Embedded Systems}{June 16--17, 2025}{Seoul, Republic of Korea}
\acmBooktitle{Proceedings of the 26th ACM SIGPLAN/SIGBED International Conference on Languages, Compilers, and Tools for Embedded Systems (LCTES '25), June 16--17, 2025, Seoul, Republic of Korea}
\received{2025-03-21}
\received[accepted]{2025-04-21}

\colorlet{yellow}{yellow!50}

\begin{document}

\title{Grouptuner: Efficient Group-Aware Compiler Auto-tuning}

\author{Bingyu Gao}
\orcid{0009-0005-3491-1384}
\affiliation{%
  \institution{MOE Key Lab of HCST (PKU), School of Computer Science, Peking University}
  \city{Beijing}
  \country{China}
}
\email{bingyugao@stu.pku.edu.cn}

\author{Mengyu Yao}
\orcid{0009-0005-8220-3470}
\affiliation{%
  \institution{MOE Key Lab of HCST (PKU), School of Computer Science, Peking University}
  \city{Beijing}
  \country{China}
}
\email{mengyuyao@stu.pku.edu.cn}

\author{Ziming Wang}
\orcid{0009-0008-1794-213X}
\affiliation{%
  \institution{MOE Key Lab of HCST (PKU), School of Computer Science, Peking University}
  \city{Beijing}
  \country{China}
}
\email{wangzim@stu.pku.edu.cn}

\author{Dong Liu}
\orcid{0009-0009-3181-5994}
\affiliation{%
  \institution{ZTE Corporation}
  \city{Chengdu}
  \country{China}
}
\email{liu.dong3@zte.com.cn}

\author{Ding Li}
\orcid{0000-0001-7558-9137}
\affiliation{%
  \institution{MOE Key Lab of HCST (PKU), School of Computer Science, Peking University}
  \city{Beijing}
  \country{China}
}
\email{ding_li@pku.edu.cn}

\author{Xiangqun Chen}
\orcid{0000-0002-7366-5906}
\affiliation{%
  \institution{MOE Key Lab of HCST (PKU), School of Computer Science, Peking University}
  \city{Beijing}
  \country{China}
}
\email{cherry@sei.pku.edu.cn}

\author{Yao Guo}
\orcid{0000-0001-5064-5286}
\authornote{Corresponding author}
\affiliation{%
  \institution{MOE Key Lab of HCST (PKU), School of Computer Science, Peking University}
  \city{Beijing}
  \country{China}
}
\email{yaoguo@pku.edu.cn}

\settopmatter{authorsperrow=4}

\begin{abstract}
Modern compilers typically provide hundreds of options to optimize program performance, but users often cannot fully leverage them due to the huge number of options. While standard optimization combinations (e.g., \textit{-O3}) provide reasonable defaults, they often fail to deliver near-peak performance across diverse programs and architectures. To address this challenge, compiler auto-tuning techniques have emerged to automate the discovery of improved option combinations. Existing techniques typically focus on identifying critical options and prioritizing them during the search to improve efficiency. However, due to limited tuning iterations, the resulting data is often sparse and noisy, making it highly challenging to accurately identify critical options. As a result, these algorithms are prone to being trapped in local optima.

To address this limitation, we propose \toolname, a group-aware auto-tuning technique that directly applies localized mutation to coherent option groups based on historically best-performing combinations, thus avoiding explicitly identifying critical options. By forgoing the need to know precisely which options are most important, \toolname maximizes the use of existing performance data, ensuring more targeted exploration. Extensive experiments demonstrate that \toolname can efficiently discover competitive option combinations, achieving an average performance improvement of 12.39\% over \textit{-O3} while requiring only 77.21\% of the time compared to the random search algorithm, significantly outperforming state-of-the-art methods.
\end{abstract}



\maketitle

\section{Introduction}
\label{sec:intro}

As a complicated software system, compiler typically provides hundreds of optimization options for users to choose from. However, effectively understanding and utilizing these options remains a significant challenge for most users. While compilers offer standard optimization combinations (e.g., \textit{-O1/-O2/-O3}), these generalized configurations often fail to deliver better performance across diverse applications and hardware architectures\cite{hoste2008cole}.

To mitigate this issue, compiler auto-tuning has emerged as a promising solution, automatically customizing optimization combinations for specific programs and hardware scenarios. Specifically, auto-tuning techniques automatically explore and identify compiler option combinations that can outperform standard configurations such as \textit{-O3}. Typically, the auto-tuning process begins by randomly generating compiler option combinations and evaluating their performance. Guided by the observed performance differences, the algorithm iteratively refines its search strategy, progressively generating and evaluating new option combinations until converging to near-optimal configurations. 

The biggest challenge in compiler auto-tuning is the sheer size of the search space, as hundreds of compiler options can lead to combinatorial explosion. Exhaustively exploring every possible combination is infeasible, and such a vast search space often causes search algorithms to become trapped in local optima. Existing techniques typically design search strategies that analyze and predict critical options based on observed data during the iterative process, and then focus the subsequent search on these critical options. For example, BOCA\cite{chen2021efficient} employs a random forest model to predict critical options during iterations, and prioritizes these critical options in subsequent searches. 

However, these methods face a critical limitation: they cannot accurately identify truly critical options. The search space of compiler auto-tuning can exceed $2^{200}$, but practical resource constraints often permit only a few hundred performance measurements. Meanwhile, each combination contains hundreds of options, while their contributions to performance are not linearly additive. Complex interactions exist among options, and not every option is valid for the program. Some options optimize features absent from the target program, yet flipping them may still mislead the search algorithm regarding truly important options. Consequently, in a sparse and noisy search space, it becomes exceedingly difficult for the tuning algorithm to accurately determine the contribution to performance of each option. Misjudging critical options can cause the tuning algorithm to either fluctuate in performance or prematurely converge by overlooking truly important options.

Compiler auto-tuning in high-dimensional and sparse space faces challenges similar to those encountered in classical high-dimensional optimization tasks. Inspired by the idea of partitioning high-dimensional variables into smaller blocks (e.g., block coordinate descent\cite{tseng2001convergence}), we argue that explicitly identifying critical options is not the only viable pathway in compiler auto-tuning. Rather than exhaustively analyzing each option in sparse and limited historical data, we can leverage historically competitive combinations and perform iterative group-level mutations. Even without knowing which specific options are crucial, their partial effectiveness has already been demonstrated in earlier evaluations. Generating new combinations based on these known-good combinations at the group-level preserves existing performance while uncovering better option combinations.

Based on this insight, we introduce \textbf{\toolname}, the first group-aware compiler auto-tuning method that moves beyond critical option reasoning and instead of focusing on functionally coherent groups based on historically best-performing combinations. \toolname features two key innovations. First, \textbf{a history-guided search process} that avoids explicit determination of critical options. Instead, it applies localized mutations to best-performing historical combinations, maximizing the use of prior performance data in a high-dimensional, sparse environment. Second, \textbf{a group-aware mutation mechanism}, in which each iteration updates only one functionally coherent group of options. By retaining the validated best-performing states of other groups, \toolname increases the likelihood of discovering beneficial interactions within the target group while maintaining overall stability in the tuning process.

We conducted extensive experiments to demonstrate the effectiveness of \toolname on three widely used benchmarks (i.e., cBench, PolyBench and SPEC CPU2017), and compared with the state-of-the-art auto-tuning algorithms, including SRTuner, BOCA and CFSCA. Experimental results show that \toolname can identify better combinations for programs in the least time. It delivers the best performance in 75.0\% of the tested programs, achieves an average 12.39\% performance improvement relative to \textit{-O3}, surpasses the best state-of-the-art method (BOCA) by 1.83\% while requiring only 39.25\% of the average tuning time of BOCA.

This paper makes the following main contributions:
\begin{itemize}
    \item We propose \toolname, the first group-aware auto-tuning algorithm that focuses on optimizing coherent option groups, significantly improving the performance of auto-tuning. The source code is available at {\hypersetup{urlcolor=black}\url{https://github.com/PKU-ASAL/GroupTuner}}.
    \item We design a history-guided search strategy that applies group-aware mutations on coherent option groups based on historically top-performing combinations to achieve incremental performance improvements.
    \item We conduct extensive experiments on three widely used benchmarks, demonstrating that \toolname achieves the best overall option combinations with the shortest tuning time, outperforming state-of-the-art auto-tuning methods. 
\end{itemize}

\section{Background}

In this section, we introduce the GCC compilation process and the optimization principles of compiler options.

The compilation process of GCC is a pipeline in which the program is gradually transformed through different intermediate representations until it is finally converted into assembly code. To organize this process efficiently, GCC employs a strategy known as \textit{pass}, which divides the entire optimization workflow into individual units. Each pass performs a specific optimization, and the output of one pass serves as the input to the next.

There are three types of passes: GIMPLE Pass, IPA Pass (Interprocedural Analysis Pass), and RTL Pass (Register Transfer Language Pass). Each type of pass operates at a different level of abstraction and focuses on optimizing specific aspects of the program. Notably, passes within the same type share similar optimization goals and characteristics. For instance, GIMPLE passes perform a series of language-independent optimizations before or after IPA passes, such as dead code elimination, which improves code structure and enhances performance. IPA passes utilize call graph information to perform transformations across function boundaries, such as function inlining, which eliminates function call overhead and improves global optimization. RTL passes finally perform target-specific low-level optimizations, such as register allocation, ensuring the generated machine code fully exploits the capabilities of the target architecture to achieve better performance. Each pass consists of two functions: \texttt{gate} and \texttt{execute}. The \texttt{gate} function determines whether the pass should be executed, while the \texttt{execute} function performs the actual optimization. The \texttt{execute} function runs only when the \texttt{gate} function returns \texttt{True}. 

\begin{figure} [t]
\centering
  \includegraphics[width=0.33\textwidth]{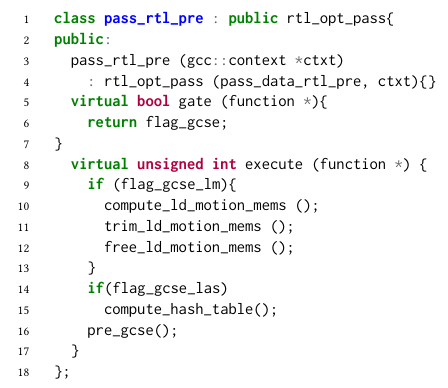}
\vspace{-0.1in}
\caption{Definition of \textit{pass\_rtl\_pre} in GCC-9.2.0.}
\label{fig:bg} 
\end{figure}

Compiler options, represented as boolean or integer variables, play a pivotal role in the \texttt{gate} and \texttt{execute} functions of passes, determining the execution of optimizations. The relationship between passes and options is many-to-many: a single pass can be influenced by multiple compiler options, and a single option can simultaneously affect multiple passes. Optimizations enabled by shared options or options within the same pass are often functionally related. For example, Figure~\ref{fig:bg} is the simplified definition of \textit{pass\_rtl\_pre} in GCC 9.2.0. This pass performs global common sub-expression elimination. The \textit{-fgcse} option controls whether this pass is enabled through its gate function. Within the execute function, the options \textit{-fgcse-lm} and \textit{-fgcse-las} further refine the optimization by specifically targeting load motion and load-after-store elimination on memory accesses.  

In this work, \toolname focuses solely on selecting whether options are enabled or disabled, rather than modifying their order, as the execution order of options is fixed and not user-configurable in GCC.

\vspace{-0.1in}
\section{ Motivation}
\label{motivation}

Most compiler auto-tuning techniques\cite{ashouri2018survey} face significant challenges in navigating the high-dimensional search space of optimization options. Existing compiler auto-tuning techniques primarily aim to identify critical options during the search process and then focus the search on these critical options to reduce the search space and improve efficiency. For instance, BOCA\cite{chen2021efficient} categorizes options into impactful and less-impactful options, and uses a random forest model to infer impactful options based on performance data collected during iterations, guiding the search to focus on impactful ones. CFSCA\cite{zhu2023compiler} first finds options potentially relevant to the target program by analyzing program structures and further identifies critical options through flip-based option selection algorithm. These critical options are subsequently prioritized in the search process to reduce the search space. Besides, SRTuner\cite{park2022srtuner} employs a multistage exploration strategy: it first identifies important options and then searches for options that synergize with them, keeping the states of these important options fixed during subsequent steps. 

\begin{figure} [t]
 \hspace*{-5mm} 
\centering
  \includegraphics[width=0.45\textwidth]{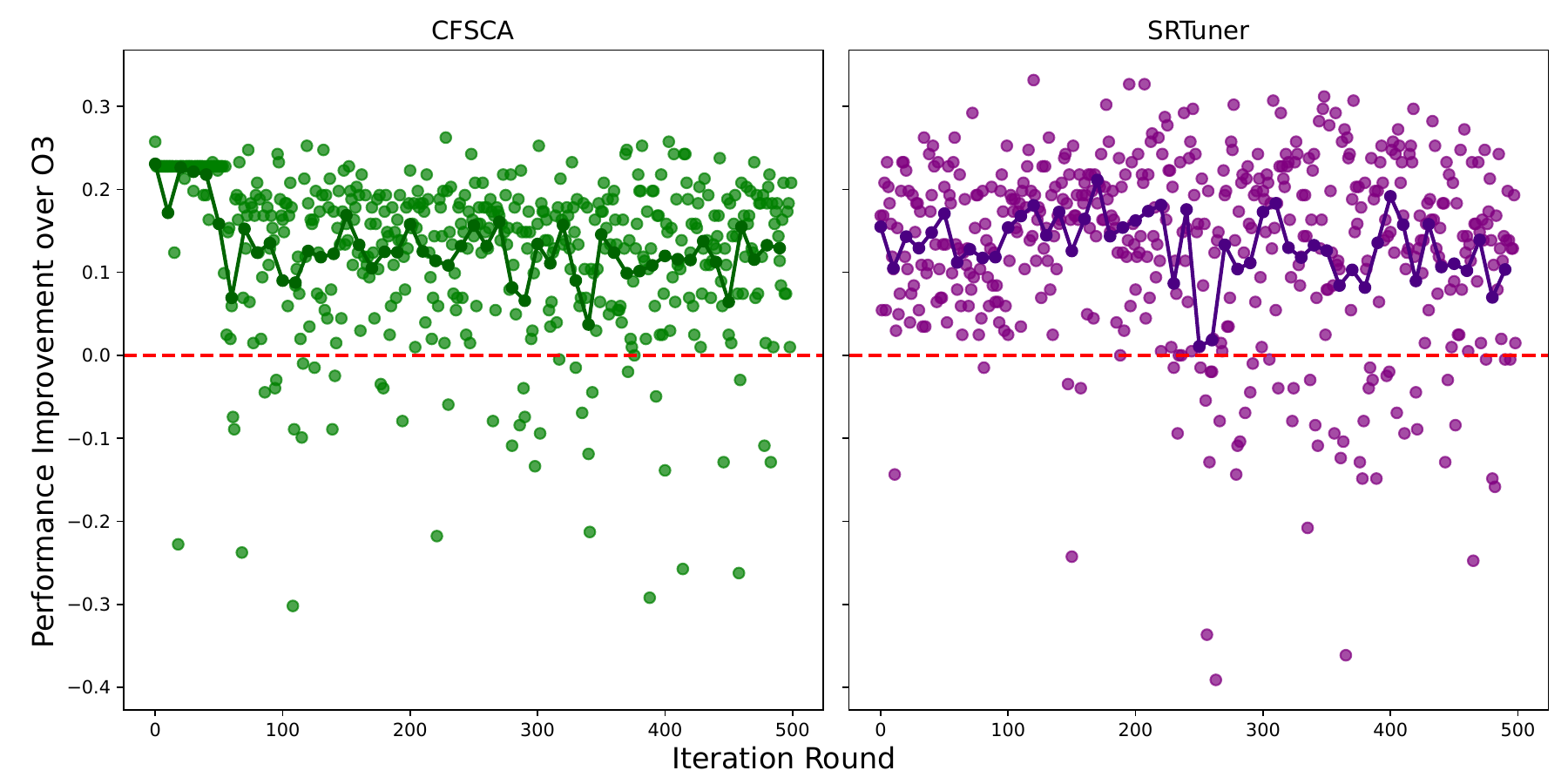}
\vspace{-0.2in}
\caption{Iterative tuning process of \textit{telecom\_adpcm\_c} in CFSCA and SRTuner.}
\label{fig:compare} 
\end{figure}

However, accurately identifying important options is extremely challenging, and incomplete or inaccurate assessments often mislead the search process causing the algorithm to converge prematurely to suboptimal combinations. On the one hand, resource constraints typically allow the auto-tuning algorithm to gather performance data only for a few hundred option combinations, whereas GCC offers over 200 optimization options, producing a search space larger than $2^{200}$. Such limited and sparse data severely hinders any attempt to accurately identify important options. On the other hand, the performance contribution of each option is far from linearly additive, as final performance typically emerges from complex interactions among dozens of options. Moreover, many options can be redundant under certain conditions, either because the program lacks the features that those options target, or because some controlling options disable them. Nevertheless, mutating these redundant options still consumes search budget and may confuse the judgments of the algorithm about which options really matter. For instance, \textit{-ftree-switch-conversion} has no effect if the program does not contain switch statements, yet toggling it can still misdirect the tuning algorithm. Finally, flipping each option individually to measure its performance is prohibitively expensive and inefficient. Figure ~\ref{fig:compare} represents the iteration process of \textit{telecom\_adpcm\_c} from cBench under CFSCA and SRTuner. The x-axis represents the iteration rounds, and the y-axis indicates the performance of the combinations generated in each round. The line represents the average performance over every ten iteration rounds, reflecting the overall performance trend of the combinations throughout the tuning process. CFSCA performs a flip-based option selection strategy to identify critical options in early phases. As shown in the left subfigure of Figure ~\ref{fig:compare}, the performance values remain nearly flat within the first 100 iterations. This suggests that due to the sparsity of truly critical options and many redundant ones, evaluating the importance by flipping individual options in a fixed combination can lead to slow progress.

Even if an algorithm manages to identify some important options, it can only confirm that these options are beneficial within a specific combination. The right subfigure of Figure ~\ref{fig:compare} illustrates the tuning process of SRTuner, which first identifies important options, and then fixes their states while further exploring options that potentially collaborate with them. However, the overall search process of SRTuner exhibits significant fluctuations, indicating that fixing the previously identified important options does not consistently lead to performance improvements when the states of their collaborating options are altered. This clearly indicates that performance improvements are not driven by single important option alone but instead rely heavily on coordinated combinations of multiple options. No single option operates in isolation, and mutating the states of other synergistic options can drastically change the overall performance.

In fact, when optimizing complex problems, researchers typically employ grouping strategies to simplify the optimization task and enhance solving efficiency. For instance, in high-dimensional optimization problems, block coordinate descent (BCD)\cite{tseng2001convergence} partitions a large set of variables into multiple blocks and optimizes each block sequentially to find local optima, iteratively repeating this process until overall convergence is achieved. 

\textbf{Overall Ideas.} Inspired by grouping methodologies, we propose that, rather than inferring critical options from limited and sparse historical data, a more reliable strategy is to directly leverage historically high-performing combinations as the starting point in each iteration. Intuitively, these evaluated high-performing combinations in earlier rounds have demonstrated partial effectiveness, offering a solid baseline from which local adjustments can uncover further gains. Since blindly modifying the states of all options globally risks destroying existing synergies and causing performance regressions, we adopt a more controlled approach, restricting modification of each iteration to a small functionally related group of options. This preserves the remainder of the combination and captures potential synergistic effects within that group. By combining historically best-performing combinations with group-level mutations, the algorithm can maintain stable, proven combinations while selectively exploring new, potentially beneficial combinations. In the following sections, we formalize this approach and present \toolname, which implements \textbf{a history-guided search} and \textbf{a group-mutation mechanism} to enhance efficiency of compiler auto-tuning.
\section{Methodology}
\begin{figure*} [t]
\hspace*{-10mm} 
\centering
  \includegraphics[width=0.95\textwidth]{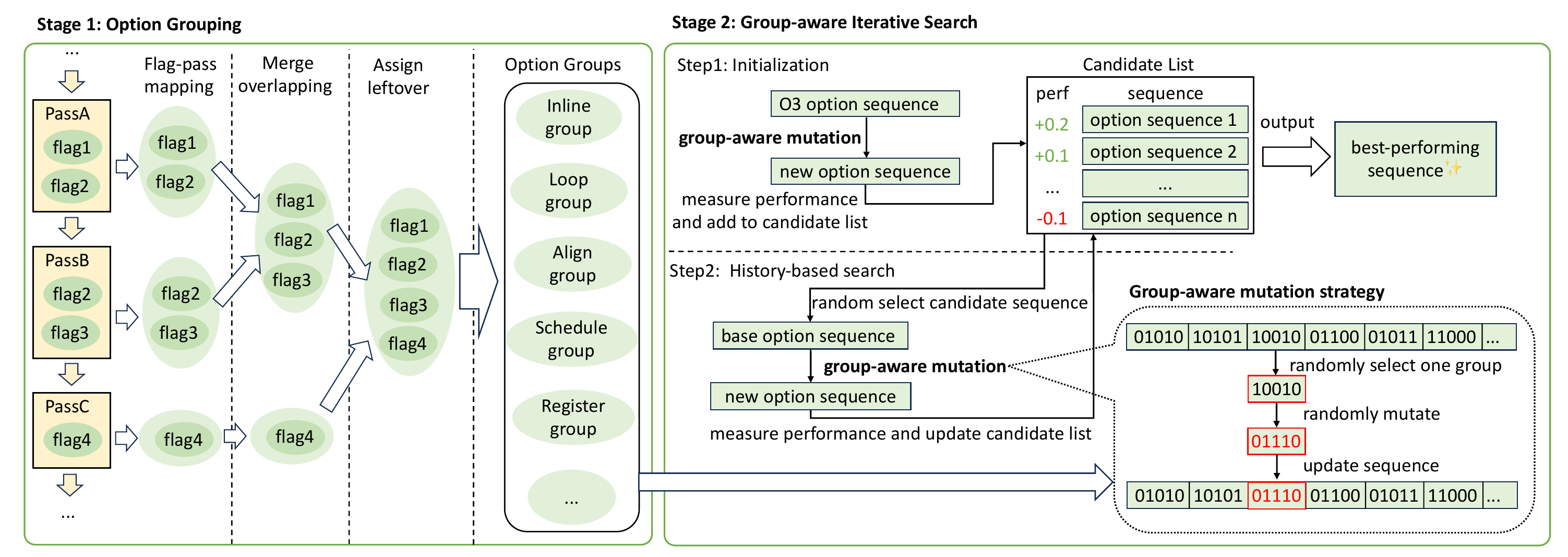}
  \vspace{-0.1in}
\caption{Overview of \toolname.}
\label{fig:overview} 
\end{figure*}

\subsection{Overview}

In this section, we introduce the detailed process of \toolname. Figure ~\ref{fig:overview} illustrates the overview of \toolname, which mainly consists of two key components: option grouping and the group-aware mutation search process. 

Specifically, option grouping first organizes compiler options into functionally coherent groups based on their relationships within passes, leveraging static analysis of compiler internals. The group-aware search process maintains a candidate list of historically top-performing combinations and applies mutations at the group level. By iteratively refining combinations based on past top-performing combinations, \toolname enables stable and incremental performance improvements. The detailed design of these two key components is described in Section ~\ref{sec:option grouping} and Section ~\ref{sec:search process}, respectively.

\subsection{Option Grouping}
\label{sec:option grouping}

For option grouping, we adopt a static grouping strategy, clustering optimization options into predefined groups. The goal of the grouping strategy is to preserve meaningful interactions among options, and reducing tuning complexity by controlling the granularity of the search space. The grouping process only needs to be done once for a specific compiler version. Although dynamic grouping may offer program-specific adaptability, it suffers from reduced accuracy and stability due to the high dimensionality of the optimization space and limited historical tuning data. Inaccurate or unstable dynamic grouping could potentially degrade tuning performance. Moreover, certain relationships among compiler optimizations inherently exist independently of specific programs or hardware architectures. For instance, alignment-related options such as \textit{-falign-loops}, \textit{-falign-jumps} and \textit{-falign-functions} naturally belong to the same group because their interactions stem from intrinsic functional characteristics rather than specific application contexts. Therefore, static grouping serves as a pragmatic and efficient strategy, minimizing tuning overhead and providing strong generality across diverse programs and architectures.

Compilers organize the optimization process into multiple stages, each consisting of several passes executed sequentially. Passes within the same optimization stage usually share similar or closely related optimization goals. Consequently, options invoked by the same pass or by passes within the same stage have a higher probability of exhibiting synergistic effects. Thus, we group options based on their association with specific optimization passes. The detailed grouping workflow consists of the following three key steps:

First, flag-pass mapping. We use CodeQL\cite{de2007keynote} to perform static analysis of the GCC source code: we first leverage control flow graph to locate functions and positions where each option is invoked, then use call graph of each pass to identify the pass associated with each option. Finally, we cluster all options belonging to the same pass into preliminary groups. 

Next, merge overlapping groups. After initial grouping, we observe overlaps among groups because certain optimizations span multiple compiler passes. Keeping related options in separate groups would increase tuning complexity and hinder capturing their inherent interactions. Therefore, we merge groups with overlapping options, ensuring closely related options are clustered into coherent groups. 

Finally, assign leftover options. After merging overlapping groups, some options remain ungrouped because they are associated with isolated passes. To avoid introducing excessive group granularity, we assign these options to the nearest existing group based on the pass execution order defined in \textit{gcc/passes.def}, under the observation that GCC structures related optimizations in close succession within the pipeline (e.g., multiple loop-related passes are grouped together).

Through this process, we categorize 206 options into 15 functionally coherent groups. As shown in Table~\ref{tab: group of options}, each group contains 4-28 options. This structured grouping significantly simplifies the tuning process, enabling targeted exploration of the optimization space. 

\begin{table*}[]
\centering
\caption{Grouping results for all optimization options in GCC. (Complete results can be found in our provided link.)}
\vspace{-0.15in}
\label{tab: group of options}
\resizebox{0.68\linewidth}{!}{
\begin{tabular}{|c|l|c|c|}
\hline
\textbf{Index} & \multicolumn{1}{c|}{\textbf{Options}}                                                                               & \textbf{Description}                                & \textbf{Size} \\ \hline
1     & -fschedule-insns -fschedule-insns2 -fsel-sched-pipelining-outer-loops -fsched-stalled-insns ...           & Instruction Scheduling              & 28      \\ \hline
2     & -fcrossjumping -fthread-jumps -fcompare-elim -fforward-propagate -fshrink-wrap-separate ...               & Branch Optimization                 & 18      \\ \hline
3     & -finline -fearly-inlining -finline-small-functions -fpartial-inlining -finline-functions ...              & Inline Optimization                 & 9       \\ \hline
4     & -ftree-pre -ftree-partial-pre -fcode-hoisting -ftree-tail-merge                                           & Redundancy Elimination Optimization & 4       \\ \hline
5     & -ftree-ccp -ftree-bit-ccp -fipa-bit-cp -fipa-cp -fipa-cp-clone -fipa-vrp                                  & Constant Propagation Optimization   & 6       \\ \hline
6     & -falign-functions -falign-jumps -falign-labels -falign-loops                                              & Alignment Optimization              & 4       \\ \hline
7     & -funroll-all-loops -fsplit-ivs-in-unroller -funroll-loops -fvariable-expansion-in-unroller -fweb ...      & Loop Optimization 1                 & 12      \\ \hline
8     & -ftree-loop-optimize -ftree-scev-cprop -fsplit-loops -fversion-loops-for-strides -ftree-ch ...            & Loop Optimization 2                 & 23      \\ \hline
9     & -fprintf-return-value -fjump-tables -ftree-slsr -fsplit-paths -foptimize-strlen -fstore-merging ...       & GIMPLE Phase Optimization 1         & 12      \\ \hline
10    & -ftree-forwprop -ftree-sra -ftree-fre -ftree-vrp -ftree-dse -fstrict-aliasing -ftree-dce -fssa-phiopt ... & GIMPLE Phase Optimization 2         & 17      \\ \hline
11    & -fdevirtualize -fipa-reference -fipa-reference-addressable -fipa-pta -fipa-icf -fipa-icf-variables ...    & IPA Phase Optimization              & 10      \\ \hline
12    & -fgcse -fgcse-lm -fgcse-las -fgcse-after-reload -fgcse-sm -fcse-follow-jumps -fdce -fdse ...              & RTL Phase Optimization 1            & 18      \\ \hline
13    & -ftree-phiprop -fstdarg-opt -fssa-backprop -ftree-builtin-call-dce -ftree-cselim -ftree-copy-prop ...     & RTL Phase Optimization 2            & 8       \\ \hline
14    & -fmerge-constants -fcx-limited-range -fcx-fortran-rules -fshort-wchar -fshort-enums ...                   & Computation Optimization            & 22      \\ \hline
15    & -ftree-ter -ftree-coalesce-vars -fdefer-pop -fconserve-stack -fsemantic-interposition -fwrapv ...         & Others                              & 15      \\ \hline
\end{tabular}
}
\end{table*}

\subsection{Group-Aware Mutation Search Process}
\label{sec:search process}
Our search process utilizes a simulated annealing-based algorithm guided by historical best observed combinations. It consists of two main phases: initialization and search process. Algorithm ~\ref{alg:optimize_algorithm} presents the detailed tuning process.

In contrast to existing methods that globally mutate multiple compiler options based on the analysis of historical data each iteration, \toolname adopts a targeted, incremental mutation strategy. Specifically, in each iteration, when generating the new combination, we first select a base combination from the candidate list that stores the best-performing combinations obtained in earlier rounds and then randomly choose one option group to mutate, altering only the states of options within this group while keeping all other options unchanged (line 2-8). Both the initialization phase and the iterative mutation phase employ this same approach. This targeted group-based mutation maximally preserves the performance benefits of the base combination while increasing the likelihood of discovering synergistic, high-impact option combinations with functionally coherent groups.

In the initialization phase, we start from the default \textit{-O3} optimization combination and randomly mutate option states at the group level to generate $n$ new diverse combinations. We measure the performance of these combinations and store them in a candidate combination list $comb_{list}$ to serve as the foundation for subsequent tuning iterations (line 10-12).

During the group mutation phase, the algorithm optimizes compiler options through the following iterative process: First, we randomly select a candidate combination from $comb_{list}$. Next, we apply a group-level mutation to generate a new combination $comb_{new}$ and measure its performance $perf_{new}$ (line 16). To balance exploration and exploitation and avoid local optima, we employ a simulated annealing-based\cite{bertsimas1993simulated} acceptance criterion (line 17-20) to update the candidate list. Specifically, if the new combination $comb_{new}$ outperforms the worst combination in the candidate list $comb_{list}$, we directly use it to replace the worst-performing combination. Otherwise, if the new combination is worse than all existing candidates, we accept it with a probability of:
\begin{equation}
P_{accept} = e^{-\Delta / (T\times \alpha)}, \quad \Delta = \frac{perf_{new} - perf_{worst}}{perf_{worst}}
\end{equation}
where $T$ is the temperature parameter initially set to $T_{0}$ and gradually reduced each iteration by $T=T \times cool_{r}$ (line 21), and $\alpha$ is a scaling factor. The iterative mutation process terminates when the temperature falls below the threshold $T<T_{min}$, and the best-performing combination from $comb_{list}$ is returned as the final result (line 22-23).

In summary, by systematically utilizing option grouping and carefully constraining mutations to functionality-related groups, \toolname ensures a targeted, incremental and efficient exploration of the compiler option space. This approach not only accelerates convergence to high-performing optimization combinations by group-level mutation but also effectively balances exploration and exploitation through the simulated annealing acceptance mechanism.

\RestyleAlgo{ruled}
\IncMargin{1em}
\LinesNumbered
\begin{algorithm}
\resizebox{0.6\linewidth}{!}{
\begin{minipage}{\linewidth}
\caption{Group-aware auto-tuning algorithm}
\label{alg:optimize_algorithm}
\SetKwInOut{Input}{input}\SetKwInOut{Output}{output}
\Input{$T_{0}, T_{min}, cool_{r}, round_{init}, Group,\alpha$}
\Output{$best_{comb}, best_{perf}$}
\SetKwFunction{GroupawareMutation}{Group\_aware\_Mutation}

\SetKwProg{Function}{Function}{:}{}

\Function{\GroupawareMutation{comb}}{
    $group \gets random\_select(Group)$\;
    \For{each $op \in group$}{
        \If{$u \sim U[0,1] \le 0.5$}{
            $group[op] \gets reverse\_state(group[op])$\;
        }
        }
    $comb_{new} \gets update(comb,group)$\;
    \Return $comb_{new}, get\_exec\_time(comb_{new})$\;
}

$comb_{list} \gets \{\} $\;
\For {$round$ from 0 to $round_{init}$}{
    $comb_{new},perf_{new} \gets group\_aware\_mutation(comb_{O3}$)\;
    $comb_{list}.add(comb_{new},perf_{new}$)\;
}

$T \gets T_{0}$\;
\While {$T > T_{min}$}{
    $comb \gets random\_select(comb_{list})$\;
    $ comb_{new},perf_{new} \gets group\_aware\_mutation(comb) $\;
    \If { $perf_{new} < perf_{worst}$}{
        $comb_{list}.replace((comb_{worst},perf_{worst}),(comb_{new},perf_{new}))$\;
         }
     \Else{
        $\Delta \gets (perf_{new}-perf_{worst})/perf_{worst} \texttt{;}$\\
       \If {$u\sim U[0, 1] < \exp(-\Delta / (T\times \alpha))$}{
          $comb_{list}.replace((comb_{worst},perf_{worst}),(comb_{new},perf_{new}))$\;
     }
     }
     $T \gets T \times cool_{r}$\;
}
$best_{comb}, best_{perf} \gets select\_best(comb_{list})$\;
\Return$(best_{comb}, best_{perf})$
\end{minipage}%
}
\end{algorithm}

\section{Evaluation}

In evaluation, we focus on the following research questions: 
\begin{tcolorbox}[size=small]
\begin{itemize}
    \item 
\textbf{RQ 1 }: Does \toolname achieve better performance compared to existing auto-tuning methods?
\item 
\textbf{RQ 2 }: Why does \toolname discover better combinations faster than existing methods?
\item 
\textbf{RQ 3 }: Does \toolname introduce additional overhead compared to existing methods?
\item
\textbf{RQ 4 }: Which option groups contribute most to performance improvements?
\end{itemize}
\end{tcolorbox}

\subsection{Experiment Setup}

\textbf{Implementation.} We implement the evaluation in Python, and our framework is extended from SRTuner\cite{park2022srtuner}. The number of iterations is set to 500. Specifically, we referred to prior works where BOCA\cite{chen2021efficient} used 120 iterations, CFSCA\cite{zhu2023compiler} operated under 6,000 seconds (approximately 100--200 iterations), and SRTuner\cite{park2022srtuner} performed 1,000 iterations. Taking these references into account, along with practical limitations in computational resources and runtime, we selected 500 iterations as a balanced compromise between tuning depth and evaluation efficiency. The size of candidate list is set to 10.
We use GCC 9.2.0 as the target compiler for x86-64 Linux platform. The initial search space comprises 206 options from 15 groups, including all options for optimizing performance collected from the GCC documentation\cite{gccOptimizeOptions} and source code\footnote{GCC defines compiler options at gcc-9.2.0/gcc/common.opt}. The experiment is conducted on a workstation running Ubuntu 20.04.6 LTS with eight 11th Gen Intel(R) Core(TM) i7-11700 @ 2.50GHz and 16GB memory. 

\textbf{Measurement.} We ensure the accuracy of the experiments in two ways: verifying program output correctness and ensuring execution time stability. For correctness, we use the output of the program compiled with \textit{-O3} as the reference standard. A result is considered valid only if the output of program matches the reference exactly. For stability, we disable the Turbo Boost feature of Intel, and run only one program at a time on the entire workstation, binding it to a fixed CPU core to prevent interference from other processes. Execution time of program is measured as the average execution time over five runs, collected using \textit{perf}\cite{perfWiki}. We use the arithmetic mean to summarize performance improvements across benchmarks, as some results are negative relative to the \textit{-O3} baseline. Since the geometric mean requires all-positive values, it is not applicable in our setting.

\textbf{Datasets.} We use three widely-used C benchmark suites: cBench\cite{cBench}, PolyBench\cite{PolyBench} and SPEC CPU2017\cite{bucek2018spec} and choose 20, 12 and 4 benchmark programs from each suite, respectively. The detailed benchmark information is shown in Table~\ref{tab:benchmark}. We did not use all available programs from these suites for three main reasons. First, certain programs produce random outputs without standard references (such as \textit{security\_sha} in cBench), making it impossible to verify the correctness of the output across different options. Second, the performance of some programs is minimally affected by options (i.e., performance gains from tuning are less than 1\%), which has been demonstrated by earlier work\cite{chen2012deconstructing}. Finally, some cases from SPEC CPU2017 require excessively long execution times, and due to resource constraints, we limited our testing to four representative cases from this suite.

\begin{table}[]
\centering
\caption{Benchmark information.}
\label{tab:benchmark}
\vspace{-0.05in}
\resizebox{0.7\linewidth}{!}{
\begin{tabular}{l|l|l!{\vrule width 1pt}l|l|l}
\toprule[1pt]
\textbf{ID} & \textbf{Program}& \textbf{Lines} & \textbf{ID} & \textbf{Program} & \textbf{Lines} \\\toprule[1pt]

C1               & automotive\_bitcount& 954        & {C19}& telecom\_adpcm\_c & 389       \\
C2               & automotive\_qsort1   & 227        &{C20}& telecom\_adpcm\_d & 391       \\
C3               & automotive\_susan\_e   & 2,129    & {P1} & cholesky     & 212       \\
C4               & automotive\_susan\_s  & 2,129     & {P2} & floyd-warshall   & 186       \\
C5               & bzip2d               & 7,200      & {P3} & gemm    & 232       \\
C6               & bzip2e                & 7,200      & {P4} & heat-3d        & 211       \\
C7               & consumer\_jpeg\_c     & 26,950     &  {P5} & lu   & 210      \\
C8               & consumer\_lame        & 21,824     & {P6} & ludcmp     & 258       \\
C9               & consumer\_tiff2bw      & 22,271     & {P7} & nussinov    & 569       \\
C10              & consumer\_tiffdither   & 22,184     & {P8}& seidel-2d   & 190       \\
C11              & network\_dijkstra      & 199        & {P9}& syr2k  & 225       \\
C12              & network\_patricia     & 634        & {P10}& symm & 231       \\
C13              & office\_rsynth        & 5,412      & {P11}& syrk & 210       \\
C14              & office\_stringsearch1    & 508        &  {P12}& trmm & 210       \\
C15              & security\_blowfish\_d& 1,524      & {S1}& mcf\_s & 3,997       \\
C16              & security\_blowfish\_e  & 1,524      & {S2}& deepsjeng\_s & 10,201      \\
C17              & telecom\_CRC32  & 307     & {S3}& imagick\_s & 263,961      \\
C18              & telecom\_gsm  & 6,049     & {S4}& nab\_s & 23,460\\

\bottomrule[1pt]

\end{tabular}
}
\end{table}
\vspace{-0.1in}

\subsection{Compared Techniques}
We compare \toolname with five algorithms, including two classic algorithms: random iterative optimization\cite{fursin2005evaluating} (RIO) and simulated annealing\cite{kirkpatrick1983optimization} (SA), and three state-of-the-art algorithms: BOCA\cite{chen2021efficient}, CFSCA\cite{zhu2023compiler} and SRTuner\cite{park2022srtuner}. RIO generates new combinations by iterating through each option in the combination and randomly selects its state. SA is configured similarly to the history-based search process of \toolname but randomly modifies option states at the global level. For the other three algorithms, we directly use the source code and the default settings they provide.

\subsection{RQ 1: Performance Improvement}
To evaluate the performance superiority of \toolname, we quantify performance improvements by comparing the best-discovered combination of each algorithm against \textit{-O3}.

Table ~\ref{tab:perf_improv} presents detailed performance improvements of different algorithms over \textit{-O3} after 500 iterations, with yellow highlighted cells indicating the best-performing algorithm for each program. \toolname achieved the best performance in 75.00\% (27/36) of programs and demonstrated the highest average performance improvement of 12.39\%, followed by BOCA at 10.56\%. SA produced the least effective tuning results, achieving an average improvement of only 9.28\%, which is even slightly lower than RIO. This limited performance primarily results from SA optimizing over the entire option combination space, significantly increasing the likelihood of altering the state of critical options within the original combination, ultimately failing to benefit from the original high-performance combination.

\begin{table}[]
\caption{Best observed performance improvements.}
\vspace{-0.1in}
\resizebox{0.60\linewidth}{!}{
\begin{tabular}{c|c|c|c|c|c|c}
\toprule[1pt]
\%  & GroupTuner     & RIO            & SA    & CFSCA          & BOCA           & SRTuner       \\\toprule[1pt]
C1  & \cellcolor{yellow}{32.23} & 30.48          & 30.36 & 31.03          & 31.32          & 30.57         \\
C2  & \cellcolor{yellow}{5.24}  & 1.13           & 1.78  & 2.29           & 2.48           & 1.21          \\
C3  & \cellcolor{yellow}{7.58}  & 4.99           & 5.24  & 4.15           & 7.36           & 4.10           \\
C4  & \cellcolor{yellow}{16.95} & 14.55          & 14.20  & 14.94          & 8.65           & 16.79         \\
C5  & \cellcolor{yellow}{3.73}  & 1.15           & 2.27  & 2.38           & 2.09           & 0.83          \\
C6  & \cellcolor{yellow}{3.53}  & 0.96           & -0.01 & -1.55          & 1.77           & 0.39          \\
C7  & \cellcolor{yellow}{21.72} & 16.20           & 20.34 & 17.72          & 17.84          & 18.52         \\
C8  & \cellcolor{yellow}{18.62} & 8.23           & 3.85  & 10.12          & 12.09          & 7.77          \\
C9  & \cellcolor{yellow}{12.64} & 10.83          & 10.99 & 12.57          & 11.15          & 10.97         \\
C10 & \cellcolor{yellow}{2.21}  & -1.45          & 0.64  & 0.64           & 0.61           & -1.21         \\
C11 & \cellcolor{yellow}{9.85}  & 1.33           & 5.09  & 5.87           & 6.77           & 1.77          \\
C12 & \cellcolor{yellow}{8.24}  & 6.58           & 6.49  & 6.11           & 6.95           & 7.11          \\
C13 & 14.60           & 9.52           & 10.99 & 14.79          & \cellcolor{yellow}{16.48} & 13.5          \\
C14 & \cellcolor{yellow}{8.50}   & 0.18           & 2.40   & 1.47           & 5.74           & 0.40           \\
C15 & 32.71          & \cellcolor{yellow}{33.66} & 30.39 & 32.01          & 31.91          & 32.23         \\
C16 & \cellcolor{yellow}{13.76} & 13.45          & 11.89 & 13.44          & 13.49          & 13.50          \\
C17 & \cellcolor{yellow}{30.29} & 29.67          & 29.49 & 29.75          & 30.09          & 29.04         \\
C18 & 26.12          & 21.85          & 19.04 & \cellcolor{yellow}{28.59} & 28.04          & 23.52         \\
C19 & \cellcolor{yellow}{35.64} & 33.17          & 31.68 & 26.24          & 31.34          & 33.17         \\
C20 & \cellcolor{yellow}{24.91} & 18.77          & 16.25 & 21.58          & 24.16          & 22.38         \\
P1  & 5.06           & 5.12           & 2.03  & 4.69           & 1.82           & \cellcolor{yellow}{6.09} \\
P2  & \cellcolor{yellow}{7.25}  & 5.00              & 6.39  & 5.64           & 4.16           & 6.06          \\
P3  & 3.25           & 2.91           & 2.47  & 2.92           & 3.46           & \cellcolor{yellow}{3.75} \\
P4  & 5.66           & 5.3            & 5.19  & 5.00              & \cellcolor{yellow}{6.17}  & 5.31          \\
P5  & \cellcolor{yellow}{3.29}  & 1.35           & 1.97  & 0.33           & 3.10           & 3.03          \\
P6  & \cellcolor{yellow}{18.34} & 17.67          & 12.23 & 16.75          & 9.16           & 14.6          \\
P7  & \cellcolor{yellow}{7.24}  & 6.27           & 6.55  & 6.54           & 6.36           & 6.96          \\
P8  & \cellcolor{yellow}{5.00}     & 4.01           & 4.62  & 4.37           & 4.42           & 4.04          \\
P9  & 7.05           & 6.92           & 5.01  & \cellcolor{yellow}{8.43}  & 3.58           & 5.38          \\
P10 & \cellcolor{yellow}{2.68}  & 1.75           & 2.19  & 1.17           & 2.32           & 2.20           \\
P11 & 8.36           & 8.08           & 8.01  & 7.78           & \cellcolor{yellow}{8.96}  & 8.31          \\
P12 & \cellcolor{yellow}{2.79}  & 2.18           & 1.74  & 2.31           & 1.71           & 1.72          \\
S1  & \cellcolor{yellow}{5.57}  & 2.39           & 3.41  & 3.36           & 4.72           & 2.82          \\
S2  & \cellcolor{yellow}{6.24}  & 3.35           & -0.93 & -2.95          & 0.53           & -4.03         \\
S3  & 8.03           & 3.33           & 3.18  & 4.68           & \cellcolor{yellow}{8.34}  & 4.07          \\
S4  & \cellcolor{yellow}{21.09} & 19.46          & 16.74 & 17.45          & 20.91          & 20.27         \\\toprule[1pt]
Avg & \cellcolor{yellow}{12.39} & 9.73           & 9.28  & 10.07          & 10.56          & 9.92         \\\bottomrule[1pt]
\end{tabular}
\label{tab:perf_improv}
}
\end{table}

\begin{figure*}
\captionsetup{font=small}
  \begin{minipage}[t]{0.3\linewidth}
    \centering
    \includegraphics[width=1\textwidth]{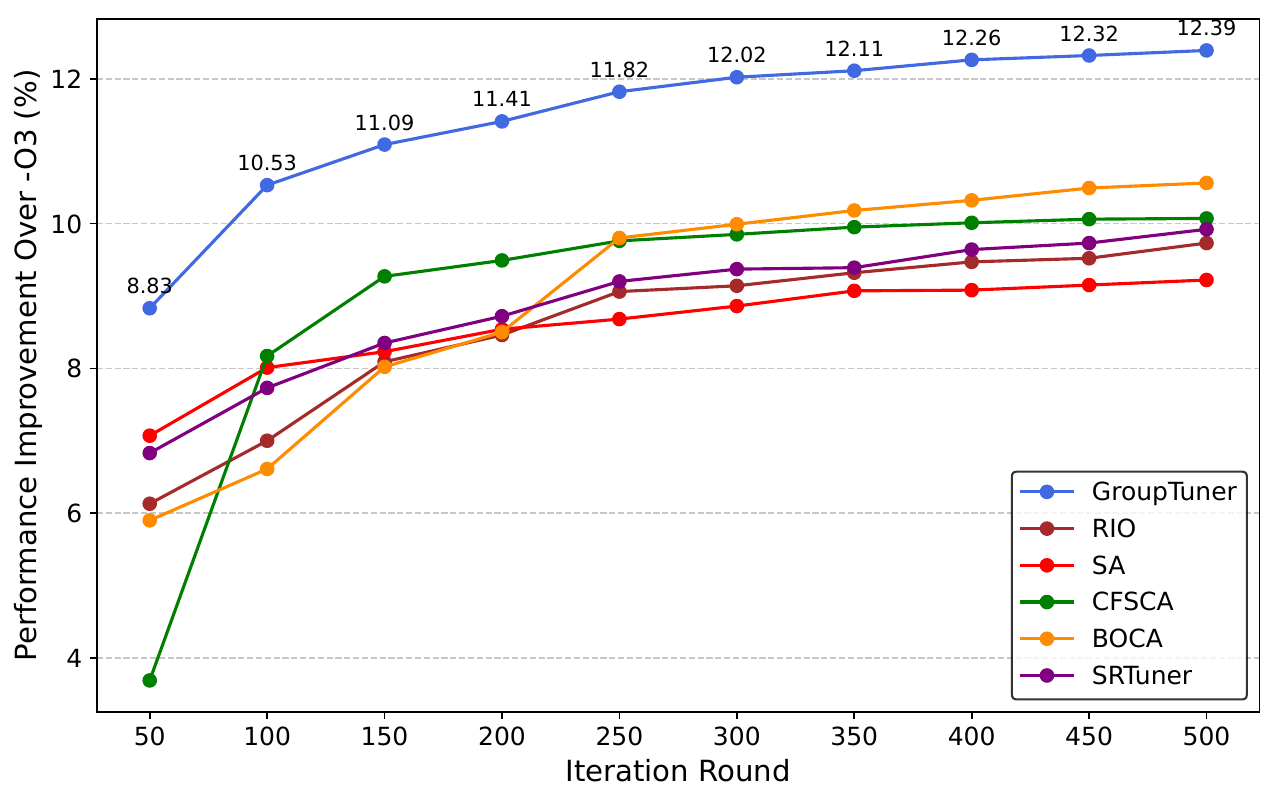}
    \caption{Performance average improvements every 50 rounds.}
    \label{fig:every 50} 
  \end{minipage}%
  \hspace{0.2em}
  \begin{minipage}[t]{0.33\linewidth}
    \centering
    \includegraphics[width=1\textwidth]{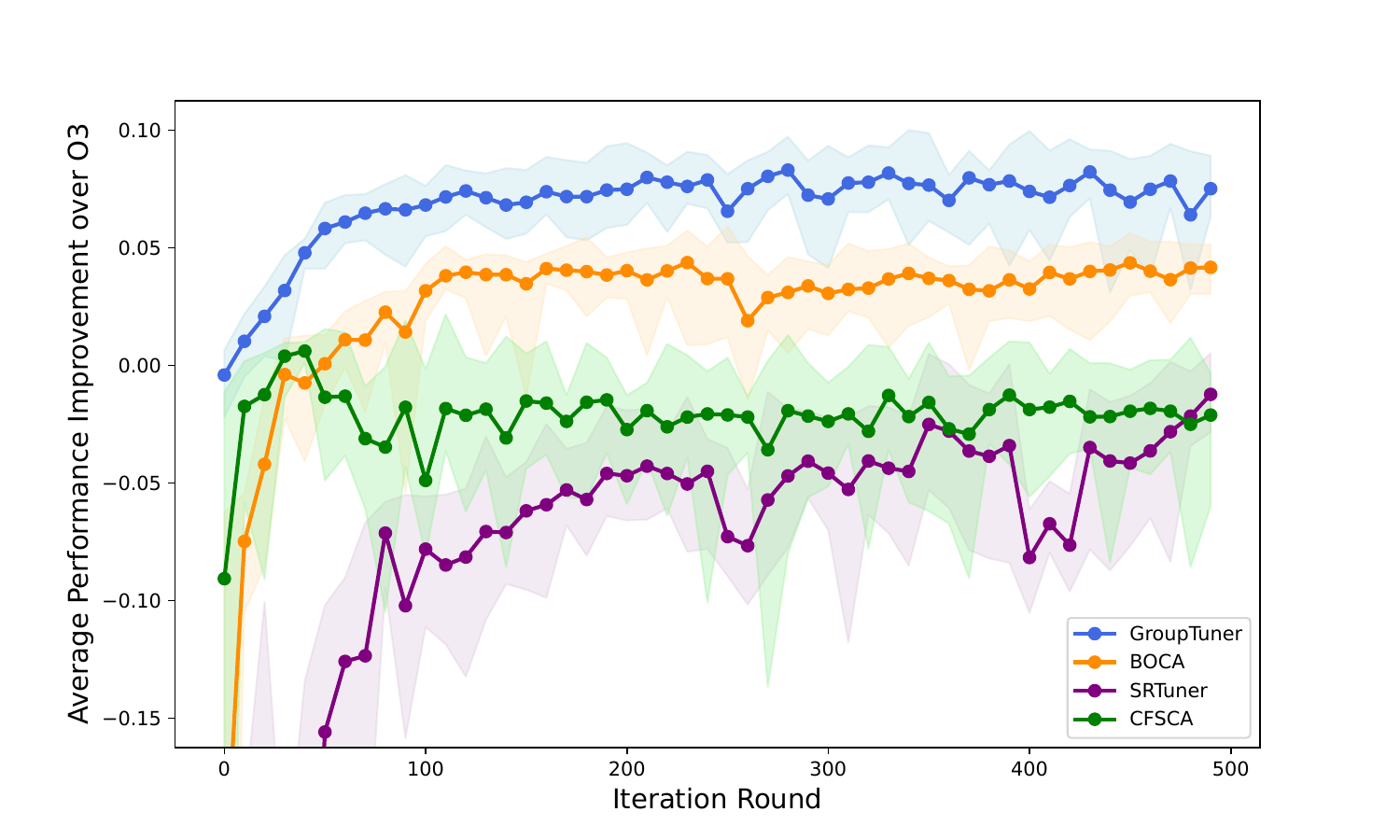}
    \caption{Average performance trend from all benchmarks.}
    \label{fig:average} 
  \end{minipage}
  \begin{minipage}[t]{0.33\linewidth}
    \centering
    \includegraphics[width=1\textwidth]{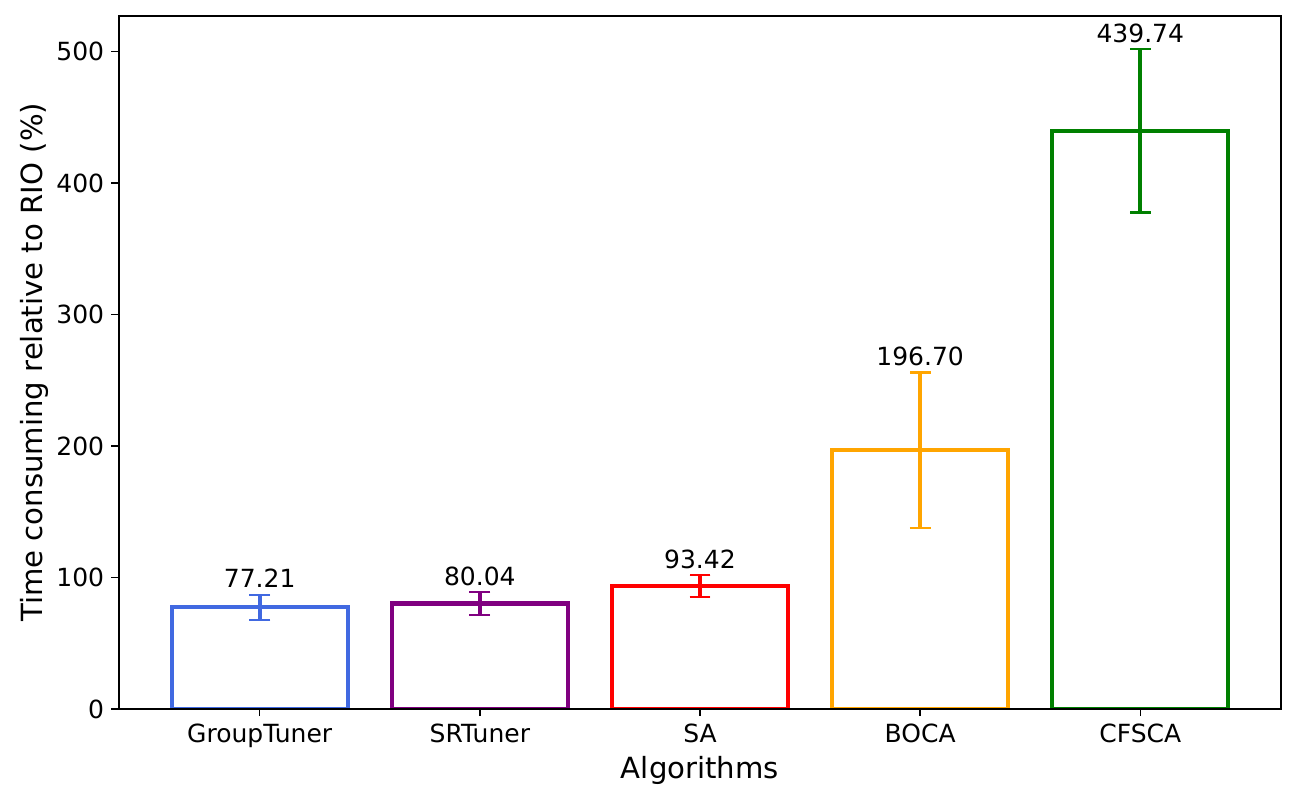}
    \caption{Average time consumption in 500 iterations of different algorithms compared to RIO.}
    \label{fig:efficiency} 
\end{minipage}
\end{figure*}

To further investigate the evolution of performance improvements during the tuning process, we track the average improvement for each algorithm at 50-round intervals and the results are shown in Figure ~\ref{fig:every 50}. As observed, \toolname consistently delivered the best performance at every iteration interval. Notably, it reached an average improvement of 8.83\% after just 50 iterations and continued discovering progressively better solutions in subsequent iterations, and achieved a 10.53\% performance improvement in just 100 iterations, nearly matching the performance gain of BOCA after 500 iterations (10.56\%). In contrast, SRTuner and SA exhibited relative good initial performance improvements (6.83\% and 7.07\%, respectively), but failed to substantially improve performance in subsequent iterations, reaching only 9.92\% and 9.28\% finally, likely due to inefficient exploration-exploitation balance. Conversely, BOCA and CFSCA algorithms showed relatively slow initial improvements (5.90\% and 3.69\%) due to their reliance on prediction models trained from historical data, which initially lacked sufficient accuracy. As iterations progressed, these models improved in accuracy, and consequently, their performance gradually improved and surpassed SA and SRTuner, achieving final average improvements of 10.56\% and 10.07\%, respectively.

\begin{tcolorbox}[size=small]
\textbf{Answer to RQ 1}: \toolname achieves the best performance on 75.0\% of programs, and gets the highest overall performance improvement at 12.39\%, up to 3.11\% better than others, and consistently maintaining the best performance throughout the entire tuning process. 
\end{tcolorbox}

\subsection{RQ2: Mechanism Analysis}

\begin{figure} [t]
 \hspace*{-5mm} 
\centering
  \includegraphics[width=0.45\textwidth]{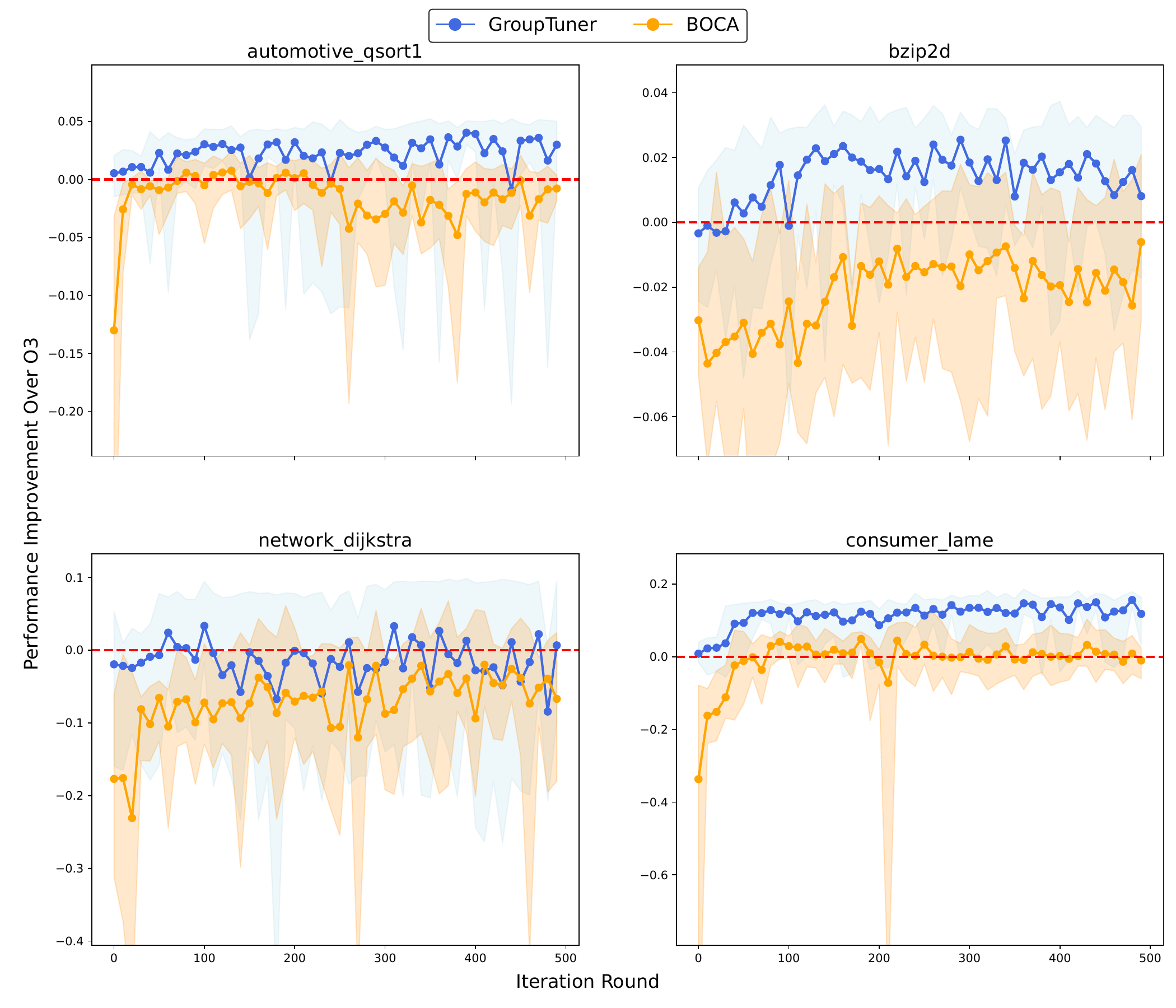}
\vspace{-0.1in}
\caption{Performance trends of option combinations found by \toolname and BOCA. }
\label{fig:trend} 
\end{figure}

To explore reasons why \toolname can effectively enhance performance, we analyzed the optimization process of some representative programs under different algorithms. Figure ~\ref{fig:trend} compares the performance improvements of combinations discovered by \toolname and BOCA during the iterative tuning process on four representative programs. Each curve shows the average performance improvement of combinations discovered every ten rounds, while the shaded area indicates the range of combination performance within those intervals, and the red horizontal line indicates the performance of \textit{-O3}. During the iteration process, \toolname gradually improves the quality of discovered combinations over time by exploring locally better combinations at the group level. Although occasional performance fluctuations may occur (e.g., as \textit{bzip2d} exhibits a temporary drop around the 100th iteration in Figure ~\ref{fig:trend}), \toolname can probabilistically discard these suboptimal combinations early on, thereby preventing them from negatively affecting the overall optimization trend. In contrast, BOCA achieves stable improvements only after accumulating sufficient historical data to accurately identify critical options. Over time, its performance gains diminish and converge prematurely, likely due to over-reliance on previously identified critical options and limited exploration of other beneficial combinations.

\begin{figure} [t]
\centering
  \includegraphics[width=0.48\textwidth]{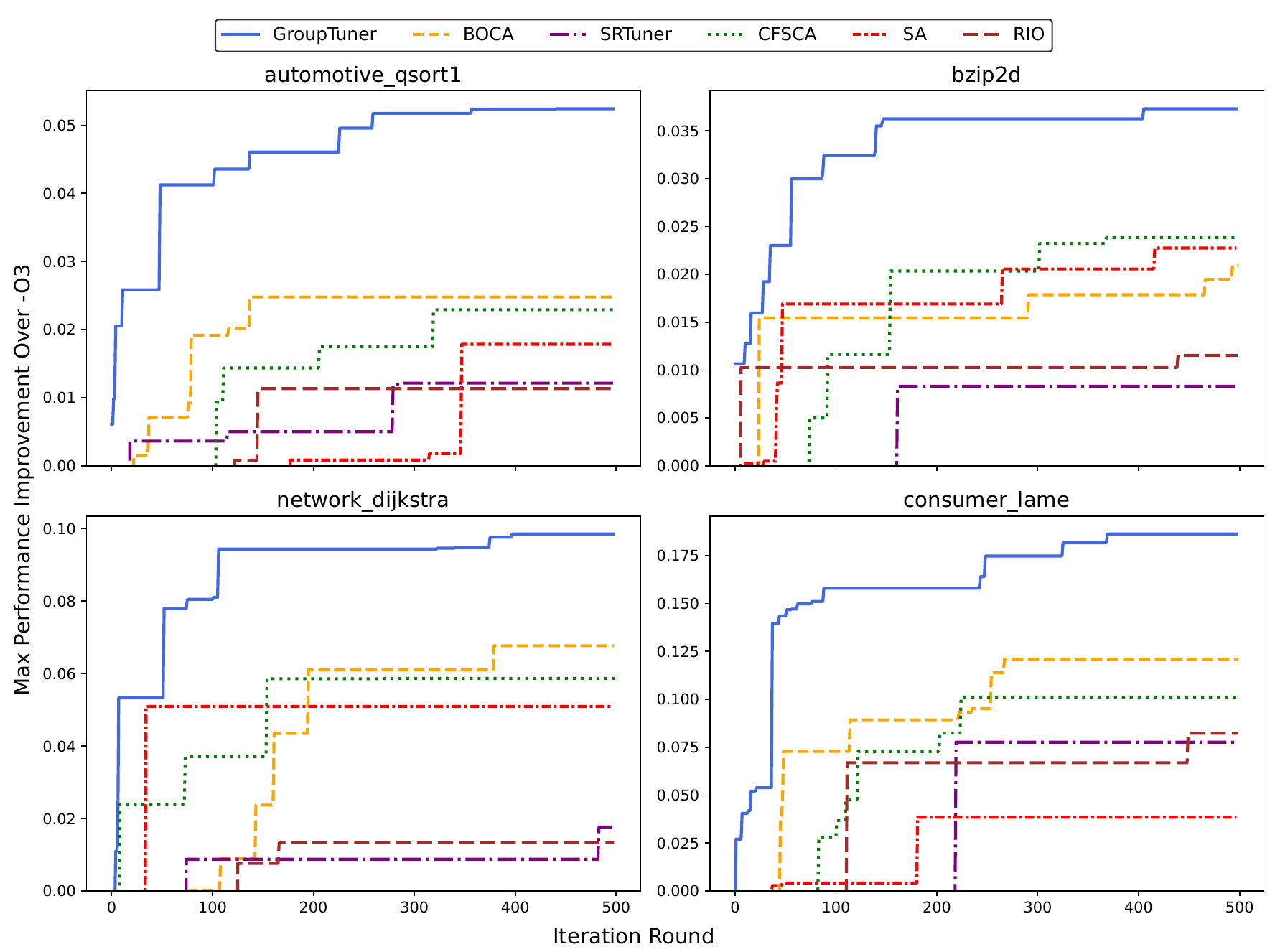}
\vspace{-0.3in}
\caption{Best observed performance improvement during tuning process.}
  \vspace{-0.1in}
  
\label{fig:max performance} 
\end{figure}

Figure ~\ref{fig:max performance} illustrates the performance evolution of best observed combinations discovered by \toolname and other algorithms for selected programs, further supporting our observations. Compared to other algorithms, combinations found by \toolname demonstrate more consistent and continuous performance improvements, highlighting the effectiveness of incremental group-level mutations over searches based solely on individual options.

To provide a broader perspective, Figure ~\ref{fig:average} illustrates the average performance improvement across all benchmarks of the Top-4 algorithms, measured every 10 iterations. The solid curves represent the average performance improvement of all benchmarks at each 10-iteration interval, while the shaded regions indicate the range of performance observed across different benchmarks during that interval. We can observe that \toolname starts with performance closer to \textit{-O3} compared to other methods. This is because \toolname initializes its tuning process by applying group-level mutation on top of the \textit{-O3} configuration, which is a high-performance setting carefully selected by compiler experts. As a result, even in the early stages, \toolname maintains a performance level near \textit{-O3}. In contrast, methods such as BOCA, SRTuner and CFSCA begin with randomly generated option combinations without leveraging \textit{-O3}, which often leads to lower initial performance. Furthermore, we can know that \toolname maintains stable improvements, systematically identifying locally improved combinations of option groups and progressively converging to global optima. 

\begin{tcolorbox}[size=small]

\textbf{Answer to RQ 2}: \toolname continuously achieves better performance improvement by fully leveraging historically top-performing combinations and performing group-aware mutation, which enable the discovery of synergistic combinations of effective options that further enhance program performance.
\end{tcolorbox}

\subsection{RQ3: Efficiency}

Efficiency is also a crucial metric for evaluating auto-tuning techniques, as the iterative nature of compilation and execution inherently incurs substantial time costs. If an auto-tuning algorithm introduces additional overhead, it risks reducing practical applicability due to excessive resource consumption. To evaluate the search efficiency, we compared the total time consumed for 500 iterations of each algorithm, defining efficiency as the time ratio relative to RIO.

Figure ~\ref{fig:efficiency} illustrates the average time relative to RIO for different algorithms. The bar represents the average tuning time across all benchmark programs in 500 iterations, while the error bar indicates the 95\% confidence interval, providing an estimate of the uncertainty in the measured time consumption. We can observe that \toolname consumes an average of only 77.21\% of the runtime required by RIO, followed by SRTuner at 80.04\%. Conversely, methods such as CFSCA and BOCA, which retrain prediction models using updated historical data in each iteration, exhibit significantly higher overheads, consuming 439.74\% and 196.70\%, respectively. This severely limits their usability. Practicality of auto-tuning is especially crucial for long-running programs, where the potential payoff from optimization is highest yet the prohibitive tuning overhead of existing methods undermines their usability. For instance, the SPEC CPU2017 benchmark \textit{mcf\_s} has a single execution time of approximately 11 seconds. \toolname completes 500 iterations in 7.61 hours, 1.47× faster than BOCA (11.15 hours) and 3.27× faster than CFSCA (24.95 hours). This efficiency gap becomes decisive in practice: for programs with hourly/daily execution times (e.g., HPC simulations), methods like CFSCA would demand weeks of tuning, whereas \toolname delivers actionable optimizations within days.

The efficiency gains of \toolname are attributed to two key factors: First, focused exploration with potentially better candidates. \toolname generates new combinations by locally flipping the state of options at the group-level based on previously identified better combinations. It ensures newly generated combinations largely remain within better regions of the search space. Even when suboptimal combinations occasionally emerge, they are probabilistically discarded early, thus avoiding prolonged execution times and unnecessary resource consumption. Second, minimal overhead due to effective use of functionality-related option groups. By leveraging functionality-related option groups as prior knowledge to guide the search process, our method incurs minimal overhead. Unlike other approaches that repeatedly train prediction models based on historical data, our method significantly reduces the iterative tuning overhead, thus substantially improving overall efficiency.

\begin{tcolorbox}[size=small]

\textbf{Answer to RQ 3}: \toolname achieves the highest efficiency, requiring only 39.25\% of the runtime required by BOCA. The group-aware mutation strategy and the history-based search process introduce almost no additional overhead during iteration.

\end{tcolorbox}

\subsection{RQ 4: Critical Groups}

\begin{figure} [t]
 \hspace*{-5mm} 
\centering
  \includegraphics[width=0.25\textwidth]{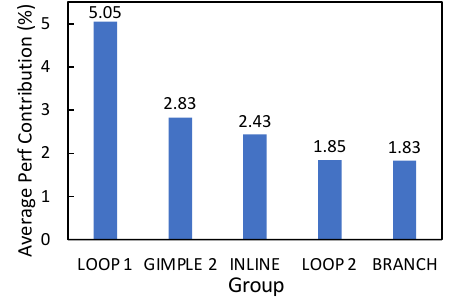}
\vspace{-0.1in}
\caption{Top-5 groups with the highest average performance contribution.  }
  \vspace{-0.2in}
\label{fig:critical groups} 
\end{figure}

To explore how \toolname effectively achieves performance gains, we analyze which option groups contribute most during tuning. We calculate the average performance improvement over \textit{-O3} brought by mutating each option group in every iteration across all benchmarks. As shown in Figure ~\ref{fig:critical groups}, the Loop 1 group achieves the highest average gain of 5.05\%, followed by GIMPLE 2 group at 2.83\%, highlighting the importance of loop and GIMPLE-phase optimizations.

To further demonstrate the necessity of group-aware tuning, we conduct a case study on the Loop 1 group, which includes options such as \textit{-ftree-loop-vectorize} and \textit{-ftree-loop-if-convert}. These options often work together: the former enables loop vectorization, while the latter simplifies control flow by converting conditional branches into branchless logic, potentially facilitating vectorization.

\begin{figure} [t]
\centering
  \includegraphics[width=0.3\textwidth]{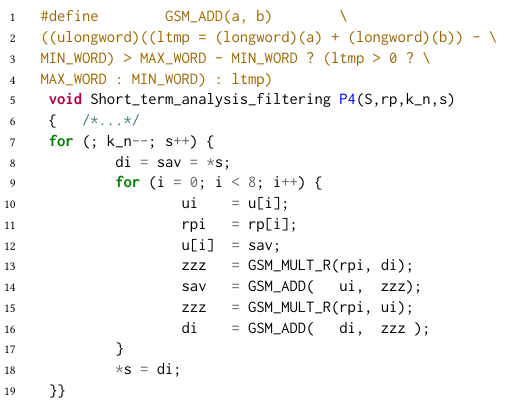}
\vspace{-0.1in}
\caption{ \textit{Short\_term\_analysis\_filtering} in \textit{telecom\_gsm}.}

  \vspace{-0.15in}
  
\label{fig:gsm}
\end{figure}

However, this synergy is not guaranteed. For instance, in \textit{telecom\_gsm} program from cBench, enabling only \textit{-ftree-loop-vectorize} has better performance than enabling both, with a maximum improvement of 11.8\% over \textit{-O3}. Further analysis reveals that its hotspot function \textit{Short\_term\_analysis\_filtering} contains an inner loop with eight iterations that repeatedly calls the \textit{GSM\_ADD} macro, which handles saturated addition with conditional logic. Figure ~\ref{fig:gsm} shows the part of logic of it. When \textit{-ftree-loop-if-convert} is enabled, these branches are transformed into branchless logic. Although this results in cleaner control flow, it increases register pressure and reduces instruction-level parallelism. Given that modern CPUs can accurately predict branches in such simple loops, the original branching structure is often faster. Moreover, due to the low iteration count and strong data dependencies within the loop, the benefits from vectorization are limited and cannot offset the overhead introduced by \textit{-ftree-loop-if-convert}, resulting in degraded overall performance.

This case demonstrates that option interactions can be either beneficial or harmful, depending on program characteristics. \toolname captures these complex interdependencies by tuning correlated options jointly within the same group, enabling more stable and robust performance improvements.

\begin{tcolorbox}[size=small]

\textbf{Answer to RQ 4}: The most critical group is LOOP 1, which achieves the average performance gain of 5.05\% during the tuning process. The contribution of each group is closely tied to the characteristics of the program.

\end{tcolorbox}
\section{Discussions}
\subsection{Limitations}
\textbf{Fixed mutation policy.} Currently, \toolname randomly selects one option group per iteration for mutation, without accounting for which groups might be more beneficial based on the characteristics of a program. For instance, programs with many function calls may gain more from inlining-related optimizations, whereas loop-intensive workloads might benefit more from loop-focused groups. As tuning progresses, an adaptive policy could increase the selection probability of highly beneficial groups and reduce the probability once their effect stabilizes. However, such dynamic strategies might also add overhead. We plan to explore more efficient adaptive group selection in future work, aiming to enhance the performance of \toolname with minimal additional cost.

\textbf{Scalability.} Our current grouping methodology is designed specifically for GCC, whose optimization structure consists of a fixed pass execution order and relies on enabling or disabling options within those passes. In contrast, LLVM exposes each optimization pass as a standalone entity and allows users to freely select, reorder, and repeat passes, greatly increasing the flexibility and complexity of the search space. Moreover, the source code of LLVM does not define explicit pass dependencies, making it challenging to directly apply our static grouping strategies. However, LLVM passes inherently exhibit implicit dependencies, even though these are not explicitly reflected in the source code of the compiler. From this perspective, the underlying principles of \toolname optimizing over functionally related groups remains applicable. To generalize \toolname to more flexible infrastructures, we envision two potential directions: (1) dynamically forming pass groups during tuning by analyzing observed dependencies or transformation interactions among passes; and (2) performing offline clustering of passes based on their impact on program IR across diverse datasets, thereby enabling the tuner to select appropriate groups based on program features. We plan to investigate these directions as part of our future work. 

Besides, traditional compilers such as GCC and LLVM determine optimization combinations based on expert knowledge, such as \textit{-O3}, which are fixed and intended to generalize across a wide range of programs. These rules are static and do not adapt based on program-specific characteristics or runtime feedback. In contrast, \toolname adopts a dynamic, feedback-driven tuning strategy that iteratively explores option space by mutating options at the group level. This allows \toolname to tailor optimizations to the specific program under tuning. While this adaptability brings performance benefits, it depends on repeated compilation and measurement, which may limit its applicability in scenarios where such feedback is unavailable or expensive to obtain.

\subsection{Threats to Validity}
\textbf{Long-Term Tuning Behavior.} Due to resource constraints, we limited the number of tuning iterations to 500 in our main experiments. However, to address concerns regarding the performance of \toolname in longer tuning durations, we conducted additional experiments on a representative subset of benchmarks, extending the tuning process to 1,000 iterations. As shown in Table~\ref{tab:perf_improv_1000}, \toolname continues to discover improved optimization combinations beyond 500 iterations. Notably, in 91.67\% (11/12) of cases, it achieved the best observed performance among all compared techniques. \toolname delivered an average 11.98\% gain over \textit{-O3}, demonstrating that our group-based strategy not only avoids premature convergence but also maintains its performance advantages during extended tuning runs.

\begin{table}[]
\caption{Performance improvements in 1,000 rounds.}
\vspace{-0.12in}
\resizebox{0.65\linewidth}{!}{
\begin{tabular}{c|c|c|c|c|c|c}
\toprule[1pt]
\%  & GroupTuner & RIO   & SA    & CFSCA & BOCA  & SRTuner \\\toprule[1pt]
C1  & \cellcolor{yellow}{47.07}      & 42.06 & 43.29 & 45.39 & 46.98 & 45.63   \\
C2  & \cellcolor{yellow}{5.35}       & 1.32  & 1.8   & 2.35  & 3.69  & 2.55    \\
C3  & \cellcolor{yellow}{11.44}      & 5.2   & 5.48  & 6.57  & 9.78  & 6.21    \\
C4  & \cellcolor{yellow}{29.9}       & 23.99 & 24.74 & 27.57 & 29.89 & 26.68   \\
C5  & \cellcolor{yellow}{4.76}       & 1.37  & 2.39  & 2.42  & 2.39  & 1.88    \\
P1  & 7.42       & 7.34  & 6.92  & 7.05  & \cellcolor{yellow}{7.85}  & 7.15    \\
P2  & \cellcolor{yellow}{8.75}       & 6.24  & 6.45  & 7.57  & 6.03  & 7.13    \\
P3  & \cellcolor{yellow}{3.95}       & 2.99  & 2.74  & 2.97  & 3.49  & 3.8     \\
P4  & \cellcolor{yellow}{7.13}       & 6.03  & 5.55  & 6.13  & 6.17  & 6.04    \\
P5  & \cellcolor{yellow}{4.85}      & 4.35  & 3.99  & 8.36  & 4.89  & 5.24    \\
S1 & \cellcolor{yellow}{6.31} & 3.33 & 4.17 & 3.70 & 4.77 & 3.26 \\
S2 & \cellcolor{yellow}{6.86} & 3.85 & 1.10 & -1.47 & 1.11 & -3.60 \\
\toprule[1pt]
Avg & \cellcolor{yellow}{11.98}      & 9.01 & 9.05 & 9.88 & 10.59 & 9.33    \\\bottomrule[1pt]
\end{tabular}
\label{tab:perf_improv_1000}
}
\end{table}

\begin{table}[]
\caption{Performance improvements in other GCC versions.}
\vspace{-0.1in}
\label{tab:other version}
\resizebox{0.55\linewidth}{!}{
\begin{tabular}{cc|c|c|c}
\multirow{2}{*}{\%} & \multicolumn{2}{c|}{GCC-11.5.0}         & \multicolumn{2}{c}{GCC-13.1.0}         \\ \cline{2-5}
                    & GroupTuner       & BOCA            & GroupTuner       & BOCA            \\ \hline
C1                  & \cellcolor{yellow}{36.67} & 35.80         & \cellcolor{yellow}{33.32} & 33.02         \\
C2                  & \cellcolor{yellow}{1.94}  & 1.08          & \cellcolor{yellow}{2.92}  & 1.90        \\
C3                  & \cellcolor{yellow}{10.85} & 8.49          & \cellcolor{yellow}{8.88}  & 6.33          \\
C4                  & \cellcolor{yellow}{25.14} & 22.80         & \cellcolor{yellow}{29.94} & 24.24         \\
C5                  & \cellcolor{yellow}{2.14}  & 0.28          & \cellcolor{yellow}{5.50}  & 4.65          \\
P1                  & \cellcolor{yellow}{5.67}  & 4.58          & \cellcolor{yellow}{5.90}  & 5.49          \\
P2                  & \cellcolor{yellow}{7.20}  & 6.18          & 4.67           & \cellcolor{yellow}{5.81} \\
P3                  & \cellcolor{yellow}{3.34}  & 3.30          & \cellcolor{yellow}{3.31}  & 3.22          \\
P4                  & 0.94           & \cellcolor{yellow}{1.11} & \cellcolor{yellow}{3.12}  & 2.52          \\
P5                  & \cellcolor{yellow}{4.41}  & 3.71          & \cellcolor{yellow}{4.03}  & 3.78          \\
S1                  &   \cellcolor{yellow}{3.73}               &     3.36            &  \cellcolor{yellow}{4.28}     &       3.54          \\
S2                  &     \cellcolor{yellow}{6.67}              &    3.06             &   \cellcolor{yellow}{0.19}                &      -2.22      \\\hline
Avg.                & \cellcolor{yellow}{9.06}  & 7.81         & \cellcolor{yellow}{8.84} & 7.69        
\end{tabular}
} 
\end{table}

\textbf{Different Versions.} Compiler version differences can also lead to performance variability. To address this concern, we performed cross-version experiments comparing \toolname with its closest competitor BOCA. These experiments involved 12 representative programs of three benchmark suites, using GCC 11.5.0 and GCC 13.1.0 under the same experimental settings. As summarized in Table~\ref{tab:other version}, \toolname achieves superior performance on 91.67\% (22/24) of test cases, delivering average improvements over \textit{-O3} of 9.06\% and 8.84\%, compared to BOCA with 7.81\% and 7.69\%. In terms of time cost, \toolname completes 500 tuning iterations in approximately 1.63 hours and 1.60 hours, whereas BOCA requires 3.53 and 3.61 hours, respectively. Overall, \toolname delivers approximately 1.16× and 1.15× the performance improvement of BOCA,  while consuming only 46.18\% and 44.32\% of the time BOCA required. These results demonstrate that \toolname enables robust performance and efficiency gains across different versions of compilers.

\section{Related Work}

Existing compiler auto-tuning techniques can be broadly categorized into three types: machine learning-based, deep learning-based and reinforcement learning-based.

\textbf{Machine learning-based and deep learning-based strategies} typically involve training a model to learn the performance variations of a program under different option combinations, gradually predicting the improved option combinations. Milepost GCC \cite{fursin2008milepost} introduced the first ML-based autotuning, relying on static features to predict effective flags. BOCA \cite{chen2021efficient} builds a Bayesian optimization model to focus on a small set of critical options, while CompTuner \cite{zhu2024compiler} uses a particle swarm optimization algorithm to refine the search based on similarity to the best-discovered combinations. In the deep learning domain, DeepTune \cite{cummins2017end} employs convolutional neural networks to learn representations directly from the intermediate representation of the program, thus reducing manual feature engineering.

\textbf{Reinforcement learning-based strategies} model the auto-tuning process as a Markov Decision Process (MDP). Park \textit{et al.}\cite{park2022srtuner} designed SRTuner, and Zhang \textit{et al.}\cite{zhang2020dynatune} designed DynaTune, both of which abstract the search strategy as a multi-armed bandit problem\cite{kuleshov2014algorithms} and use UCB\cite{garivier2011upper} (Upper Confidence Bound) to balance exploration and exploitation. CompilerGym\cite{cummins2022compilergym} is the first reinforcement learning environment specially designed for the compiler optimization task. Trofin \textit{et al.} proposed MLGO\cite{trofin2021mlgo}, which uses policy gradient algorithms and evolutionary algorithms to train decision networks through reinforcement learning for better inlining optimization decisions. 

Additionally, recent work has begun exploring the use of large language models for compilation optimization\cite{cummins2024meta,grubisic2024compiler}, including optimizing code size\cite{cummins2023large}, loop vectorization\cite{taneja2024llm}, function calls\cite{singh2024llm}, etc. 
\section{Conclusion}

In this paper, we present \toolname, a group-aware compiler auto-tuning that avoids explicitly identifying critical options in a high-dimensional and sparse search space from limited historical performance data during iteration. By leveraging historically best-performing combinations and applying localized mutation at the granularity of functionally coherent option groups, \toolname maximizes the use of limited performance data while substantially reducing the risk of disruptive global changes. Extensive experiments show that \toolname consistently discovers improved compiler option combinations with minimal time overhead, outperforming other state-of-the-art methods.

\section*{Data-Availability Statement}

The artifact of this paper is available at Zenodo: \cite{our-artifact}.

\section*{Acknowledgments}

We thank the anonymous reviewers for their comments. This work was partly supported by the National Key R\&D of China (2022YFB4501802), Beijing Natural Science Foundation (L243010) and ZTE Corporation.

\onecolumn
\begin{multicols}{2}
\bibliographystyle{ACM-Reference-Format}
\bibliography{software}
\end{multicols}

\end{document}